\documentclass[11pt,a4paper]{article}
\usepackage{jheppub}
\usepackage{setspace,braket}
\usepackage{amsmath, amssymb, bm}
\usepackage{etoolbox}

    \makeatletter
    \patchcmd{\maketitle}{\@fpheader}{}{}{}
\makeatother
\def\be{\begin{equation}}
\def\ee{\end{equation}}

\def\({\left(}
\def\){\right)}
\def\[{\left[}
\def\]{\right]}

\def\lm{\lambda}

\newcommand{\bea}{\begin{eqnarray}}
\newcommand{\eea}{\end{eqnarray}}

\def\d#1#2{\frac{\displaystyle #1}{\displaystyle #2}}

\numberwithin{equation}{section}
\renewcommand{\thesection}{\arabic{section}}

\begin{document}
\renewcommand{\thefootnote}{\fnsymbol{footnote}}

\title{Barrow black holes and the minimal length}
\author[a,b]{Li-Hua Wang,}
\author[a,b]{Meng-Sen Ma\footnote{Corresponding author. E-mail address: mengsenma@gmail.com}}
\affiliation[a]{Department of Physics, Shanxi Datong
University,  Datong 037009, China}
\affiliation[b]{Institute of Theoretical Physics, Shanxi Datong
University, Datong 037009, China}

\abstract{
Following Barrow's idea of fractal black hole horizon, we re-derive black hole entropy of static spherically symmetric black holes. When a black hole absorbs matter its horizon area will increase. Given the spherically fractal structure, we conjecture that the minimal increase of the horizon area should be the area of the smallest bubble sphere.  From this, we find the black hole entropy has a logarithmic form, which is similar to that of Boltzmann entropy if we consider $A/A_{pl}$ as the number of microscopic states. We further calculate temperatures and heat capacities of Schwarzschild, Reissner-Nordstr{\"o}m(RN), and RN-AdS black holes. It is found that their temperatures are all monotonically increasing and the heat capacities are all positive, which means these black holes are thermodynamically stable. Besides, for RN-AdS black hole we find its heat capacity has Schottky anomaly-like behavior, which may reflect the existence of the discrete energy level and restricted microscopical degree of freedom.

}
\maketitle
\onehalfspace

\renewcommand{\thefootnote}{\arabic{footnote}}
\setcounter{footnote}{0}
\section{Introduction}
\label{intro}

Bekenstein and Hawking found that in the framework of general relativity(GR) black hole entropy is proportional to the horizon area of a black hole, to be precise, it is  $S=A/4$. In other theories of gravity, the Bekenstein-Hawking entropy can be modified by contributions from higher-order curvature terms and can be simply derived using the Wald formula. Even in GR, when some quantum effects are taken into account, the area law of black hole entropy can also be corrected. 

The most commonly considered quantum effect is the generalized uncertainty principle (GUP).  GUP predicts the existence of a minimal length scale of the order of the Planck length, which can also be deduced from string theory and other tentative theories of quantum gravity\cite{Konishi.276.1990,Maggiore.65.1993,GARAY.145.1995,Kempf.7909.1997,Scardigli.39.1999,Capozziello.15.2000}. There are more than one expressions for GUP, of which the most simple form is
\be
\Delta x \geq \d{\hbar}{\Delta p}+\d{\alpha^2}{\hbar}\Delta p \geq 2\alpha \sim l_p,
\ee
where $l_p$ is the Planck length and $\alpha$ is a positive constant. With the correction of the GUP, black hole thermodynamics can be significantly changed\cite{Adler.2101.2001,Medved.124021.2004,Nozari.156.2006,Ren.208.2006,Myung.393.2007,Nouicer.63.2007,Kim.035.2008,Park.698.2008,Xiang.046.2009,Ma.861.2014,Tawfik.1430025.2014,Miao.1.2015,Feng.276.2016,Vagenas.40001.2017}. In general, black holes no longer evaporate completely, but leave behind a remnant at finite temperature. The black hole entropy will receive a logarithmic correction term proportional to $\ln A$ besides the leading Bekenstein-Hawking entropy $A/4$.

The microscopic mechanisms behind gravitation and black hole entropy are yet to be fully understood. Recently, Barrow proposed a toy model for possible effects of quantum gravity by considering the fractal structure of horizon surface\cite{Barrow.2020.135643}. In this case, the area and volume of a black hole should  be  the sum of all the intricate structures. With these fractal structures, the entropies of a black hole and our Universe can be very large. Barrow's model has also been further extended to the study of dark energy\cite{Saridakis.102.123525,Moradpour.80.732} and black hole thermodynamics\cite{Abreu.2020, Abreu.2020.135805}. 

It is our concern in this letter that how the existence of a minimal length will affect the entropy of black holes with fractal structures. In this case, when a black hole absorbs a particle there will be a natural minimal increase in the horizon area, $\Delta A_{min}$. Following the idea of \cite{Medved.124021.2004}, we can derive the black hole entropy. According to the first law of black hole thermodynamics, we can further derive the temperature and heat capacity. We find that for black holes with fractal horizon these thermodynamic quantities are much different from that of standard black holes.

The paper is arranged as follows. We first simply review Barrow's fractal black hole and some of his results in section 1. In section 2 we re-derive black hole entropy based on this fractal structure. We then calculate the temperature and heat capacity of some black holes with fractal structure. At last, we summarize our results and discuss the possible future study.

\section{Barrow black holes with fractal structure}

In this part, we give a brief introduction about Barrow's idea on the fractal structure of the black hole horizon. For details, one can refer to Barrow's paper.
Barrow considered Schwarzschild black hole and imagined there are many smaller spheres attaching on the black hole horizon and then more smaller spheres attached to these spheres and so on. This is a fractal structure.

Suppose that at each step there are $N$ spheres and the radius is $\lambda$ times smaller than that of the sphere in the previous step.  Let $r_0=r_h$, which is the Schwarzschild radius. If there is no cut off at some small finite scale, the actual surface area of the horizon and volume of the black hole are infinite series:

\be
A_{\infty}=\sum_{n=0}^{\infty} N^{n} 4 \pi\left(\lambda^{n} r_{h}\right)^{2}=4 \pi r_{h}^{2} \sum_{n=0}^{\infty}\left(N \lambda^{2}\right)^{n},
\ee

\be
V_{\infty}=\sum_{n=0}^{\infty} N^{n} \frac{4 \pi}{3}\left(\lambda^{n} r_{h}\right)^{3}=\frac{4 \pi r_{h}^{3}}{3} \sum_{n=0}^{\infty}\left(N \lambda^{3}\right)^{n}.
\ee

Barrow discussed that when $\lambda^{-2}<N<\lambda^{-3}$ the surface area will be infinite and the volume of the black hole is finite. 

Besides, on the basis of the area law of black hole entropy Barrow further conjectured that the entropy can take the form of $S \approx A/A_{pl} \approx (A_h/A_{pl})^{(2+\Delta)/2}$, where $A$ and $A_h$ are the area of the fractal horizon and the standard horizon, $A_{pl}$ is the Planck area. 
$0<\Delta<1$ with $\Delta=0$ corresponding to the standard horizon and $\Delta=1$ corresponding to the most intricate horizon.

\section{The Entropy of Barrow black holes}

In quantum gravity there is the idea of quantized spactime, which means the existence of the smallest finite length scale. Usually the Planck length $l_p$ is considered to be this scale. The GUP, which gives an apparent minimal length, is a realization of this idea. With this cut-off, Barrow black hole can have interesting thermodynamic properties.

Although $l_p$ puts the lower limit of length scale, the radius of the smallest sphere in the fractal structure need not exactly to be $l_p$. In fact, according to the recurrence relation $r_{n+1}=\lambda r_n$ it only needs, at some cut-off step $n_1$, to satisfy
\be
r_{n_1} \geq l_p, \quad  \lambda r_{n_1} < l_p.
\ee

In this case, the surface area should be 
\be\label{actual_area}
A=4 \pi r_{h}^{2} \sum_{n=0}^{n_1}\left(N \lambda^{2}\right)^{n}=A_h\d{1-(N\lm^2)^{n_1+1}}{1-N\lm^2},
\ee
where $A_h=4\pi r_h^2$ is the original area of black hole horizon.

Next we try to derive the black hole entropy by following the approach of the work \cite{Medved.124021.2004}.
\be\label{SA}
\frac{d S}{d A} \simeq \frac{(\Delta S)_{\min }}{(\Delta A)_{\min }},
\ee
where  $(\Delta S)_{\min }$ represents the minimal increase of entropy, the value of which is a constant $\ln2$ according to information theory. In the following  we mainly focus on the calculation of the minimal increase of the surface area, $(\Delta A)_{\min }$.

Considering the fractal structure, when the black hole absorbs a  particle the minimal increase of total area should be the area of the smallest sphere, namely
\be\label{deltaA}
(\Delta A)_{\min }=4\pi r_{n_1}^2=4\pi (r_h\lm^{n_1})^2=\lm^{2n_1} A_{h}.
\ee

Combing Eqs.(\ref{actual_area}) and (\ref{deltaA}), one can obtain
\be\\\\\\\\\\\\\\\label{deltaA2}
(\Delta A)_{\min }=Ac_1,
\ee
where we have set $c_1= \d{\lm^{2n_1}(1-N\lm^2)}{1-(N\lm^2)^{n_1+1}}$, which should be a positive constant.

Substituting Eq.(\ref{deltaA2}) into Eq.(\ref{SA}), we can obtain
\be
\frac{d S}{d A} \simeq \frac{(\Delta S)_{\min }}{(\Delta A)_{\min }}=\d{\ln 2}{c_1A}.
\ee

Clearly, we have
\be\label{Barrow_entr}
S=\d{\ln 2}{c_1} \ln\d{A}{A_{pl}} +c_0,
\ee
where we introduce  $A_{pl}$ to obtain a dimensionless quantity in the logarithmic function. Below we will call this result as Barrow entropy for short.

Clearly, the Barrow entropy does not satisfy the usual area law, or the logarithmic correction to area law. However, this form makes one reminiscent of the well-known Boltzmann formula: $S=k_B\ln\Omega$. If we make the correspondence $A/A_{pl} \leftrightarrow \Omega$ and calibrate $\ln 2/c_1$ to one\footnote{Because we take the natural unit. The Boltzmann constant $k_B$ has been set 1.} and set the integration constant to be zero, we can understand the Barrow entropy as Boltzmann entropy. In analogy to statistical mechanics, we can consider $A_{pl}$ as the area occupied by one microscopic state and therefore $A/A_{pl}$ is just the number of microscopic states of the black holes, $\Omega$.

\section{Temperature and heat capacity of Barrow black holes}

As a thermodynamic system, the thermodynamic quantities of black holes should satisfy the thermodynamic identity:
\be\label{1stlaw}
d M=T d S+ \cdots.
\ee

On the basis of Barrow entropy, below we will further analyze the influence of the fractal structure on the thermodynamic properties of several black holes.

\subsection{Schwarzschild black hole}

If the metric of Schwarzschild black hole is not affected, we can derive the temperature 
\be\label{SchT}
\d{1}{T_{B}}=\d{\partial S}{\partial M}=\d{4\ln 2}{c_1}\d{1}{r_h},
\ee
where the subscript ``B" is for Barrow for short.

Clearly, this temperature is proportional to the Schwarzschild radius, while in standard Schwarzschild black hole the Hawking temperature is inversely proportional to $r_h$. Until now, we still do not know how black hole temperature depends on horizon radius through observation.  We cannot directly rule out this possibility. Moreover, with this temperature, we find that the heat capacity of Schwarzschild black hole is always positive:
\be
C_{B}=\d{\partial M}{\partial T} = \d{2\ln 2}{c_1}>0.
\ee
Therefore, with this fractal structure Schwarzschild black hole can be thermodynamically stable.

\subsection{Reissner-Nordstrom(RN) black hole}

Barrow's idea can also be used for other static spherically symmetric black holes. We first take RN black hole as an example.
The line element is
\be
ds^{2}=-f(r)dt^{2}+f(r)^{-1}dr^{2}+r^{2}d\Omega_2^2,
\ee
with the the metric function
\begin{equation}
	f(r)=1-\d{2M}{r}+\d{Q^2}{r^2},
\end{equation}%
where the parameters $M,~Q$ are the ADM mass and electric charge of the black hole.

The position of the event horizon of the black hole can be determined as the larger root of $f(r_h)=0$. Reversely, one can express the black hole
mass $M$ according to $r_h$,
\be
M=\frac{Q^2+r_h^2}{2 r_h}.
\ee

The temperature corresponding to the Bekenstein-Hawking entropy is
\be
T=\d{f'(r_h)}{4\pi}=\d{r_h^2-Q^2}{4\pi r_h^3}.
\ee

We then calculate the heat capacity at constant $Q$, which is a counterpart of the heat capacity at constant volume, $C_V$, in $PVT$ system.
\be
C=\left.\d{\partial M}{\partial T}\right|_Q=\frac{2 \pi  r_h^2 \left(Q^2-r_h^2 \right)}{r_h^2-3 Q^2}.
\ee

Using the Barrow entropy and similar to Eq.(\ref{SchT}), we can obtain the Barrow temperature of RN black hole
\be
T_{B}=\frac{c_1 \left(r_h^2-Q^2\right)}{ r_h 4\ln 2}.
\ee

Furthermore, the heat capacity should be
\be
C_{B}=\left.\d{\partial M}{\partial T_B}\right|_Q=\frac{\ln 4 \left(r_h^2-Q^2\right)}{c_1 \left(r_h^2+Q^2\right)}.
\ee

\begin{figure}
	\centering{
	\includegraphics[width=7cm]{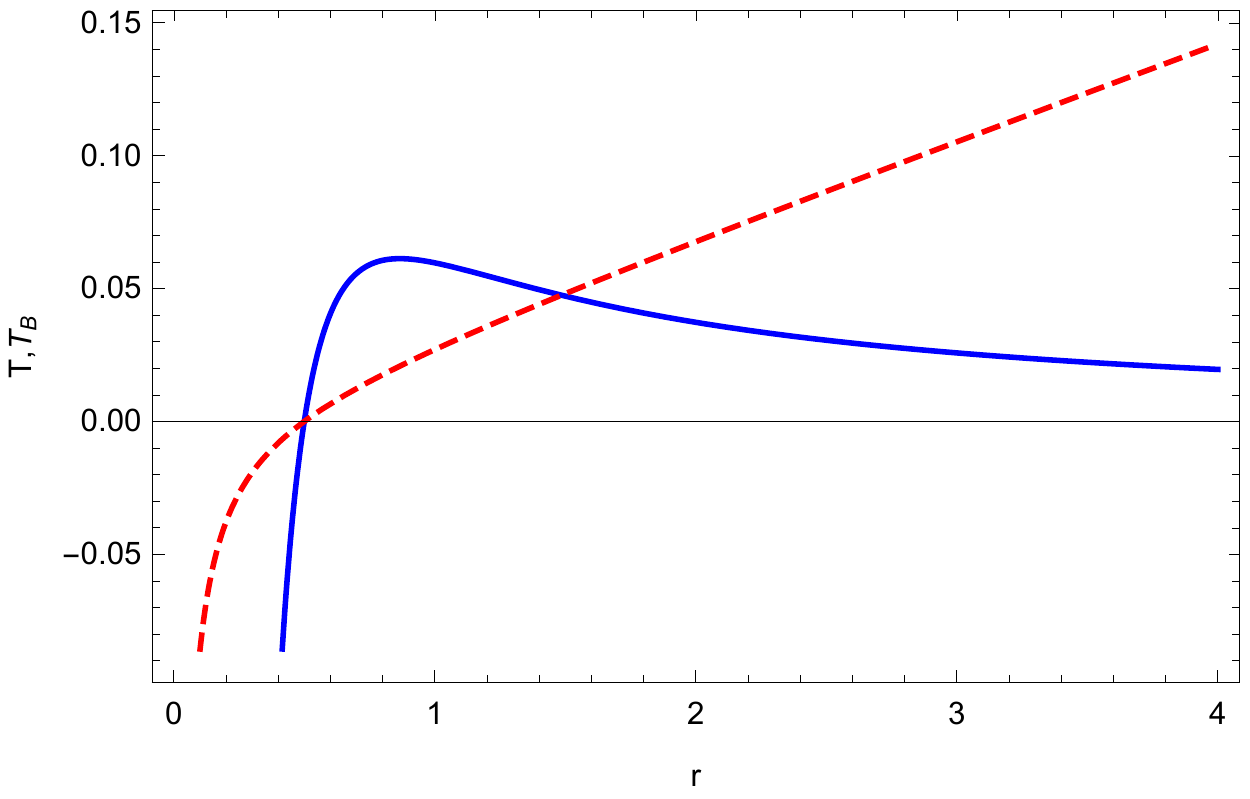} 
	\caption{The Barrow temperature $T_B$ (the red dashed line) and Hawking temperature $T$ (the blue solid line) for RN black hole. We have set $Q=0.5,~ c_1=0.1$.} \label{fig_tem}
	}
\end{figure}

\begin{figure}
	\centering{
	\includegraphics[width=7cm]{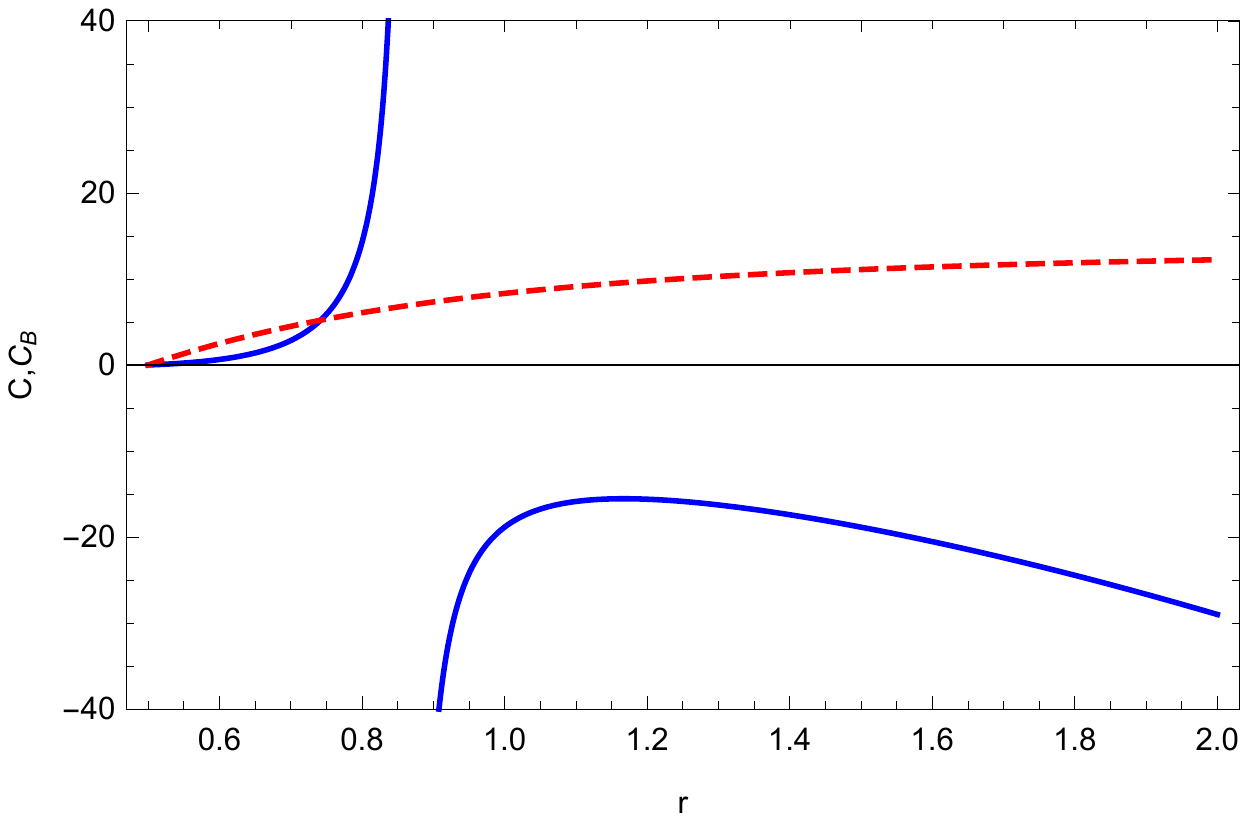}
	\caption{$C_B$  (the red dashed line) and $C$ (the blue solid line) are the heat capacities of the RN black hole with fractal structure and the standard RN black hole, respectively. We have set $Q=0.5,~ c_1=0.1$.}
	\label{fig_C}}
\end{figure}

From Fig.\ref{fig_tem}, we can see that the two temperatures both become zero in the extremal limit, $r_h=Q$. As $r_h$ increases, the Barrow temperature is monotonically increasing, while the Hawking temperature first increases to a maximum and then decreases monotonically. Correspondingly, as is shown in Fig.\ref{fig_C}, the heat capacity for the RN black hole with fractal structure is always positive and finite, while for the standard RN black hole the heat capacity is positive only at a finite interval and turns negative after undergoing a divergent point.  

\subsection{RN-AdS black hole}

For RN-AdS black hole, the metric function is
\be
f(r)=1-\d{2M}{r}+\d{Q^2}{r^2}+\d{r^2}{l^2},
\ee
where $l$ represents the cosmological radius.

The mass function can be expressed according to the horizon radius,
\be
M=\frac{l^2 Q^2+l^2 r_h^2+r_h^4}{2 l^2 r_h}.
\ee
The standard Hawking temperature is
\be
T=\frac{l^2 r_h^2+3 r_h^4-l^2 Q^2}{4 \pi  l^2 r_h^3}.
\ee
And the heat capacity  is
\be
C=\frac{2 \pi  r_h^2 \left(l^2 r_h^2+3 r_h^4-l^2 Q^2\right)}{-l^2 r_h^2+3 r_h^4+3 l^2 Q^2}.
\ee

\begin{figure}
	\centering{
	\includegraphics[width=7cm]{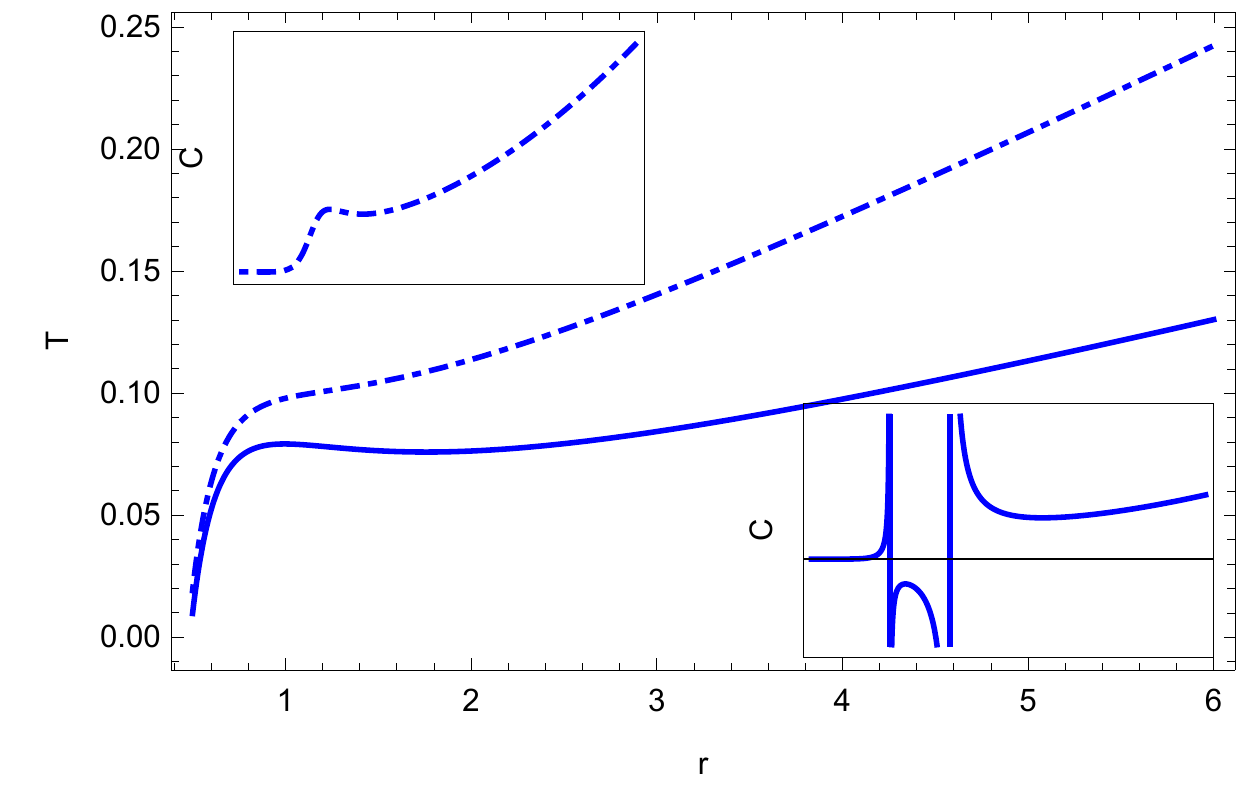}
	\caption{The  Hawking temperature of RN-AdS black hole  as a function of horizon radius. The solid and dot-dashed curves correspond to $l=4$ and $l=2.5$, respectively. Besides, we set $Q=0.5,~ c_1=0.1$. We also sketch the behaviors of heat capacities at different values of $l$.} \label{figTAdS}
}
\end{figure}

As is shown in Fig.\ref{figTAdS}, for different values of $l$ this temperature can increase monotonically or have some extrema. In the former case, the heat capacity is always positive. In the latter case, the heat capacity is positive only in the interval where the slope of the temperature is positive. This configuration corresponds to the first-order phase transition between the smaller black hole and the larger one.

Similar to the RN black hole, we can also obtain the Barrow temperature and the corresponding heat capacity, which are
\bea
T_B&=&\frac{c_1 \left(l^2 r_h^2+3 r_h^4-l^2 Q^2\right)}{4\ln 2 l^2  r_h}, \\
C_B&=&\frac{2\ln2 \left(l^2 r_h^2+3 r_h^4-l^2 Q^2\right)}{c_1 \left(l^2 r_h^2+9 r_h^4+l^2 Q^2\right)}.
\eea

\begin{figure}
	\centering{
	\includegraphics[width=7cm]{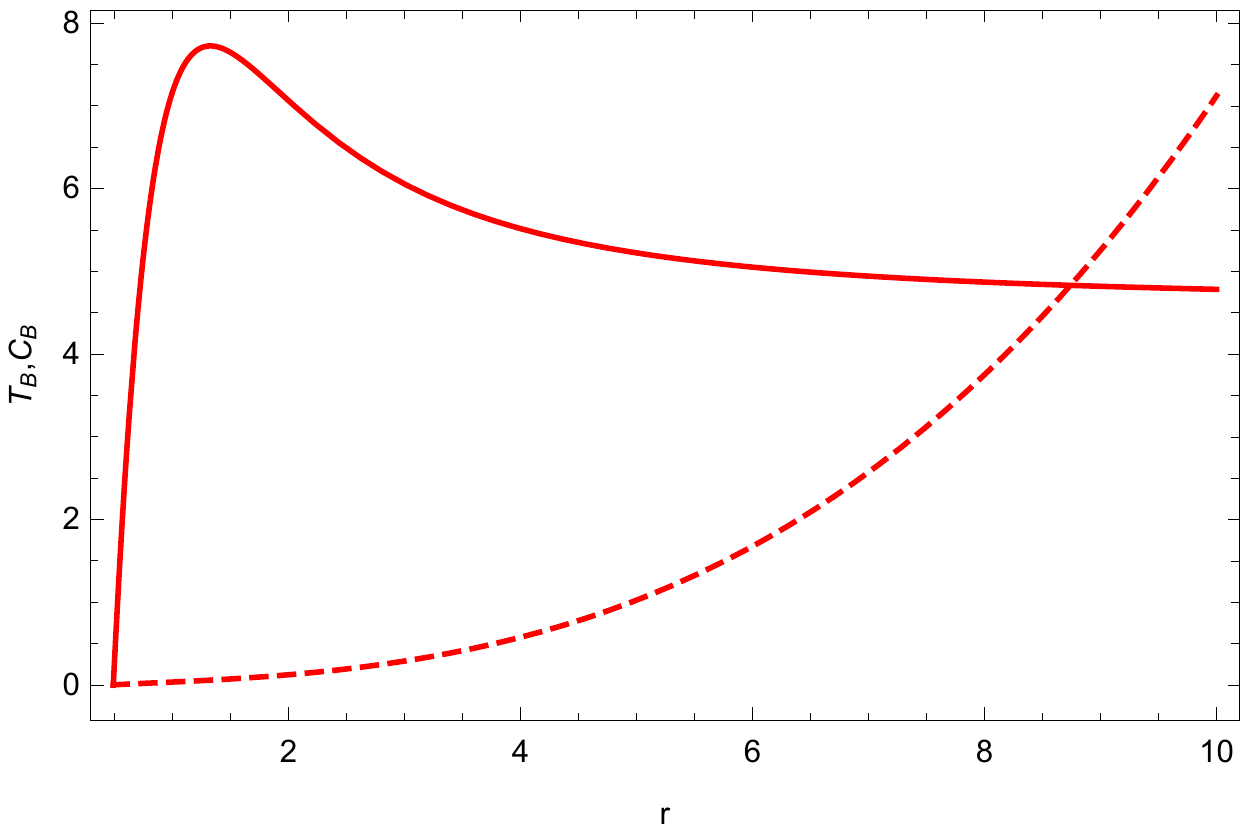}
	\caption{The Barrow temperature (the red dashed curve) and heat capacity (the red solid curve) of the RN-AdS black hole with fractal structure. We have set $Q=0.5,~ c_1=0.1,~ l=4$.}\label{figTBAdS}}
\end{figure}

Their behaviors are depicted in Fig.\ref{figTBAdS}. Similar to Schwarzschild and RN black holes, the Barrow temperature is also monotonically increasing. The heat capacity  is always positive for positive  Barrow temperature. 

\begin{figure}
	\centering{
	\includegraphics[width=7cm]{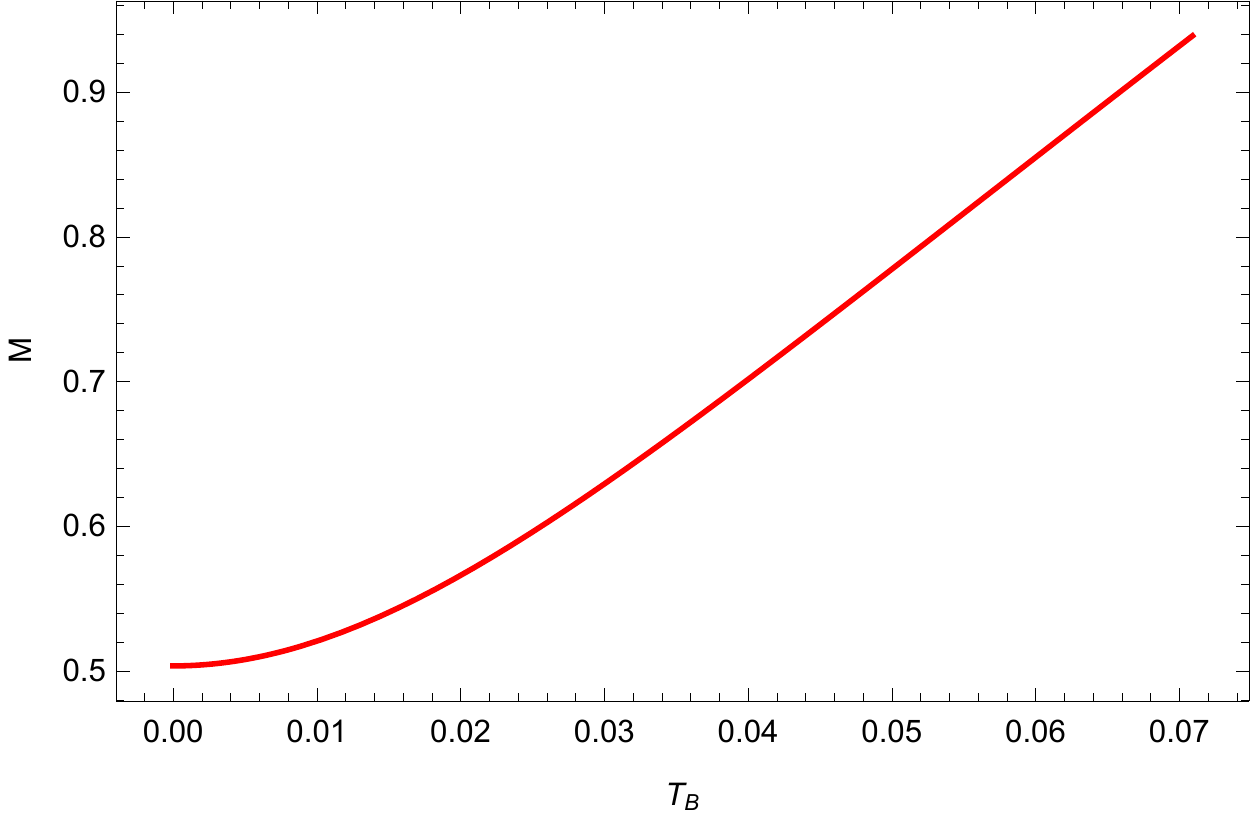}\hspace{0.5cm}
	\includegraphics[width=7cm]{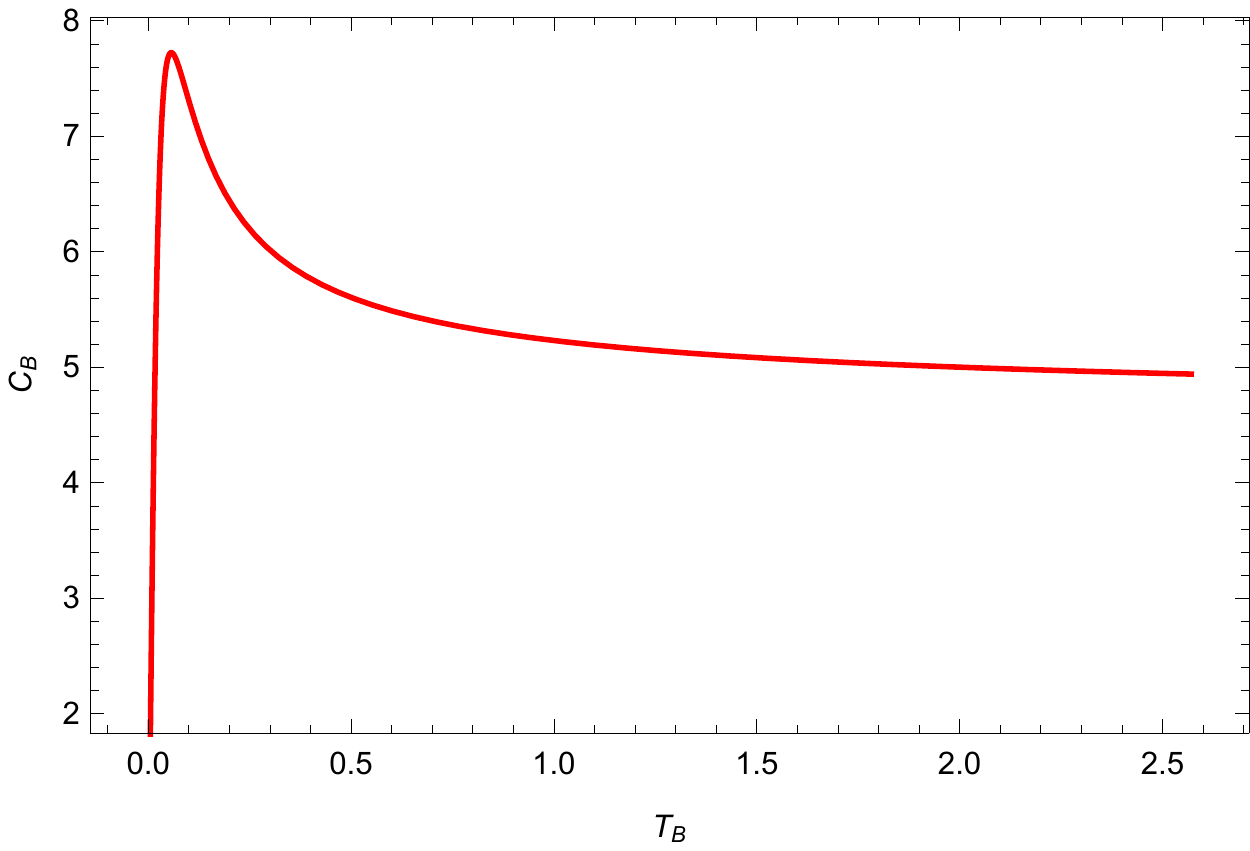}  
	\caption{The mass and the heat capacity as a function of Barrow temperature for the RN-AdS black hole with fractal structure. We have set $Q=0.5,~ c_1=0.1,~ l=4$.}\label{schottky}}
\end{figure}

To further clarify the relation between $C_B$ and $T_B$, we depict the curve of $C_B-T_B$ in Fig.\ref{schottky}.
$C_B$ tends to zero as $T_B \rightarrow 0$ and tends to a constant when $T_B \rightarrow \infty$, and in between it has a maximum. This behavior of $C_B$ has a striking resemblance to the Schottky anomaly of heat capacity of ordinary solid matter. The existence of the peak means the RN-AdS black hole should have discrete energy levels microscopically. If we treat the black hole mass $M$ as the internal energy, then $M$ should be the statistical average of these energy levels.  $M$ increases monotonically with $T_B$, but at an ever increasing rate. At high enough temperature, $M$ is nearly proportional to $T_B$. And the heat capacity keeps a constant value thereafter.

\section{Discussions and Conclusions}
\label{Conclusions}

On the basis of Barrow's idea of the fractal horizon and the existence of a minimal length, we try to re-derive black hole entropy. The result relies on the choice of the minimal increase of horizon area, $\Delta A_{min}$. Due to the fractal structure, we think that the most natural choice should be the area of the smallest bubble sphere. In this way, we found that the entropy has a logarithmic form for static spherically symmetric black holes. This Boltzmann-like entropy may reflect the microscopic structure of black holes if we consider $A/A_{pl}$ as the number of microscopic states.

We assumed the laws of black hole thermodynamics still hold and further calculated the temperature and heat capacity of these black holes with fractal structures. First, the temperature is generally given by
\be
T_B=\d{\partial M}{\partial S_B}=\d{\partial M}{\partial S}\d{\partial S}{\partial S_B}=T\d{\partial S}{\partial S_B}=\frac{\pi  c_1 }{\ln 2}r_h^2 T.
\ee
On the one hand, this result guarantees that the Barrow temperature has the same sign as that of the Hawking temperature. On the other hand, the factor $r_h^2$ makes the Barrow temperature increase more quickly for large $r_h$ and leads to a monotonically increasing $T_B$. Therefore, for the black holes we considered,
\be
C_B=\d{\partial M}{\partial T_B}=T_B\d{\partial S_B}{\partial T_B}=T_B\d{\partial S_B/\partial r_h}{\partial T_B/\partial r_h}>0.
\ee
This means these black holes are,  at least locally, thermodynamically stable. Besides, the monotonicity of $T_B$ means that $\partial T_B/\partial r_h \neq 0$,  which implies that the heat capacities are continuous and have no divergent points. This indicates that the phase structures of these black holes are very simple.

There is an unexpected result for RN-AdS black hole with fractal structure. Its heat capacity exhibits a Schottky anomaly-like behavior, which has also been found and discussed in other gravitational system\cite{Grumiller.044032.2014,Johnson.054003.2020,Dinsmore.054001.2020}. This can be attributed to the existence of discrete energy levels and restricted microscopic degrees of freedom.  At least there are some low-lying energy levels separated from the remainder of the energy spectrum. At very low temperatures, the heat capacity increases rapidly with the temperature. At high enough temperature, the behavior of any thermodynamic system approaches that of its classical counterpart. In this case, $k_BT$ is much larger than the interval of adjacent energy levels, so the energy levels are quasi-continuous. In thermodynamic systems with finite energy, such as two-level system and dS black holes\cite{Dinsmore.054001.2020}, the heat capacity should tend to zero in the high temperature limit. But the heat capacity of RN-AdS black hole with fractal structures has a nonzero value in this limit. This reveals that the total mass $M$ of this black hole has no upper limit. As the temperature increases, the black hole can always absorb heat and become more energetic.   

 The presence of the discrete energy levels must be relevant to the fractal structure of the RN-AdS black hole. However, the Schwarzschild black hole and the RN black hole do not possess this property. We also want to know whether these interesting thermodynamic properties also exist for other black holes.  It is of great interest to extend our current study to the higher-dimensional and more complicated spherically symmetric black holes, which may provide new insights toward a better understanding of the microscopic structure of black holes.

\bigskip
\bigskip

\textbf{Declaration of competing interest}

The authors declare that they have no known competing financial interests or personal relationships that could have appeared to
influence the work reported in this paper.

\acknowledgments
This work is supported in part by Shanxi Provincial Natural Science Foundation of China (Grant No. 201701D121002), by Scientific and Technological Innovation Programs of Higher Education Institutions in Shanxi (Grant No. 2021L386) and by Datong City Key Project of Research and Development of Industry of China (Grant No. 2018021).

\bibliographystyle{JHEP}
\bibliography{E:/mms/References/criticalGUP}

\end{document}